# Denoising the 3-Base Periodicity Walks of DNA Sequences in Gene Finding

Changchuan Yin*
College of Natural Sciences, University of Phoenix Chicago Campus, IL 60173, USA
Email: cyinbox@email.phoenix.edu

Dongchul Yoo [1] and Stephen S.-T. Yau [2]
[1] Department of Mathematics, Statistics and Computer Science, University of Illinois at Chicago, IL 60607, USA
[2] Department of Mathematical Sciences, Tsinghua University, Beijing 100080, China
Email: dcyoo@math.uic.edu，syau@math.tsinghua.edu.cn

*Abstract*—A nonlinear Tracking-Differentiator is one-input-two-output system that can generate smooth approximation of measured signals and get the derivatives of the signals. The nonlinear tracking-Differentiator is explored to denoise and generate the derivatives of the walks of the 3-periodicity of DNA sequences. An improved algorithm for gene finding is presented using the nonlinear Tracking-Differentiator. The gene finding algorithm employs the 3-base periodicity of coding region. The 3-base periodicity DNA walks are denoised and tracked using the nonlinear Tracking-Differentiator. Case studies demonstrate that the nonlinear Tracking-Differentiator is an effective method to improve the accuracy of the gene finding algorithm.

## I. INTRODUCTION

Identifying protein-coding regions of eukaryote genomes is one of the most fundamental tasks for genome analysis, but poses important challenges as protein-coding regions (exons) are usually flanked by non-coding regions (introns) in eukaryote genomes. There are no distinguished sequence features between exon and intron sequences. Despite a variety of statistics methods developed to predict exons such as GenScan [1] and MZFF [2], many issues remain unsolved, for example, short exons are still difficult to be identified correctly [3].

During last decades, a number of approaches have been suggested for differentiating between the protein-coding and non-protein-coding regions of DNA, Fourier transform time-frequency analysis has attracted significant attentions in the identification of exons. This method is based on the 3-base periodicity [4], [5], measured as the Fourier power spectrum of a DNA sequence at the frequency 1/3. It was well-known that the 3-base periodicity exits in many exon sequences but not in intron sequences due to the non-uniform distribution of the four nucleotides *A, C, T, G* in protein-coding regions and uniform distribution in the non-coding regions [6], [7]. Derivatives of the 3-periodicity walks along DNA sequences are employed in the gene finding algorithm, however, the trajectories are not smooth, it is difficult to compute directly the derivatives of 3-base periodicity walks to determine exons from DNA sequences.

A nonlinear Tracking-Differentiator, proposed by Han and Wang [8], [9], is one-input-two-output system that can generate smooth approximation of measured signals and get the derivatives of the signals. Many theoretical aspects of this method have been studied to improve and extend the method during last decade [10], [11]. It has been applied in various applications such as motion control and online estimator [12], [13].

In this paper, we apply the nonlinear Tracking-Differentiator to smooth the walks of the 3-base periodicity of DNA sequences to develop an effective gene finding algorithm. Case studies show that the tracking method described in this paper is an effective way to design a gene prediction algorithm in terms of accuracy. This method may also hold prospective promises in other data processing fields of Bioinformatics.

## II. SYSTEM AND METHODS

### A. Nonlinear Tracking-Differentiator

Han and Wang [8, 9] proposed a nonlinear Tracking-Differentiator, which generates continuous and differential signals properly from those that are not differentiable. The nonlinear Tracking-Differentiator is a parameterized second-order system with one input and two outputs. Let $v_1(t)$ denote the input of the tracking differentiator, and $x_1(t), x_2(t)$ as the two outputs, where $x_1(t)$ tracks $v_1(t)$, and $x_2(t)$ is the derivative of $x_1(t)$. In general, the state space representation is used to describe the system of a linear TD. The following section illustrates the theory of the nonlinear TD [8],[9].

*Theorem 2.1* If the system

is globally finite time stable at (0, 0), then for any bounded integrable function $v(t)$ and $T>0$, with the setting time $T(x_0)$. The solution of the following system:

$$\begin{cases} \dot{x}_1 = x_2 \\ \dot{x}_2 = R^2 f(x_1 - v(t), \frac{x_2}{R}) \end{cases}$$

satisfies

$$\lim_{R \to \infty} \int_0^T |x_1(t) - v(t)| dt = 0$$

Where $R$ represents the maximum acceleration that the system can reach and is a function of the maximum actuation available in the system.

The function $f(x_1, x_2)$ can be chosen as a nonlinear function. Two nonlinear functions are described as follow:

$$f(x_1, x_2) = -|x_1|^{\alpha_1} \operatorname{sgn}(x_1) - |x_2|^{\alpha_2} \operatorname{sgn}(x_2)$$

$$f(x_1, x_2) = -\operatorname{sgn}(x_1 + \frac{1}{2}|x_1|x_2)$$

where $\operatorname{sgn}(.)$ denotes sign function. By using the synthetic function based on the Isochronic region method, Han and Yuan [13] suggested the discrete form of TD to complement its performance as follows:

$$\begin{cases} X_1[\lfloor t/h \rfloor + 1] = X_1[\lfloor t/h \rfloor] + h X_2[\lfloor t/h \rfloor] \\ X_2[\lfloor t/h \rfloor + 1] = X_2[\lfloor t/h \rfloor] - h fst(X_1[\lfloor t/h \rfloor] \\ \qquad - V[\lfloor t/h \rfloor], X_2[\lfloor t/h \rfloor], R, h) \end{cases}$$

where $h$ is a sampling step size, $V[\ ]$ is the input signal, $X_1[]$ and $X_2[]$ are the two output signals from TD, and $R$ is an acceleration factor. The function $fst(X_1, X_2, R, h)$ is defined as follows:

$$fst(X_1, X_2, R, h) = \begin{cases} -R(a/d), |a| \le d \\ -R \operatorname{sgn}(a), |a| > d \end{cases}$$

where
$$d = Rh,$$
$$d_0 = dh,$$
$$y = vX_1 + hX_2,$$
$$a_0 = \sqrt{d^2 + 8R|y|},$$
$$a = \begin{cases} X_2 + y/h, |y| < d_0 \\ X_2 + \operatorname{sgn}(y)(a_0 - d)/2, |y| \ge d_0 \end{cases}$$

The nonlinear TD is a noise filter, which blocks any part of the signal with the acceleration exceeding $R$, providing an efficient way to generate smooth approximations. Moreover, nonlinear TD is able to obtain the derivatives of a signal with noise that is too noisy to differentiate directly. The following is a numerical simulation for the nonlinear TD using the following input signal:

$$v = x^2 + 10 * random$$

where random is a *random* function with mean is 0. Fig. 1 illustrates the input signal and its smooth approximation by the nonlinear TD.

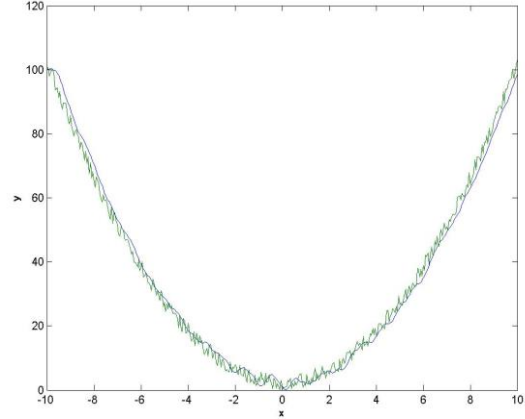

Figure 1. Smoothing the test function by Nonlinear TD with R = 0.5.

### B. Computing the 3-base periodicity of a DNA sequence

Number The 3-base periodicity is computed from the nucleotide distributions [14]. Let $F_{x1}, F_{x2}, F_{x3}$ be the occurrence frequencies of the nucleotide $x \in \{A,T,C,G\}$ in the first, the second and the third codon position, respectively. The 3-base periodicity, *PS(N/3)*, is calculated as follows:

$$\sum_{x=A,T,C,G} [F_{x1}^2 + F_{x2}^2 + F_{x3}^2 - (F_{x1}*F_{x2} + F_{x2}*F_{x3} + F_{x1}*F_{x3})]$$

The Fourier spectrum background is determined by the length of the DNA sequence [6]. The ratio of the 3-base periodicity signal to spectrum background of a DNA sequence can be interpreted as the normalized strength of the 3-base periodicity per base pair of the DNA sequence.

### C. Generation of the 3-base periodicity walks of DNA sequences

We design the following algorithm to generate the trajectories of 3-base periodicity of DNA sequences. For a DNA sequence of length *N*, let $D_k$ denote the DNA walk sequence of length k, *i.e.*, $D_k$ is the sub-region of the DNA sequence ranging from the beginning to the position k. The algorithm is described as follows:

1. Set *k = 1*.
2. Compute the 3-base periodicity *PS(N/3)* in $D_k$ based on the nucleotide distributions in the three codon positions of $D_k$, $F_{xi} \in \{A,T,C,G\}, i \in \{1,2,3\}$. The nucleotide distribution of a DNA walk of length *k* can be obtained recursively from the DNA walk sequence of length *k-1* with the occurring frequencies of the nucleotides on the position *k*.

3. Increase k by 1 and repeat step 1 to step 2, until *k = N*.

### D. Predicting exons by tracking the trajectories of 3-base periodicity

The 3-base periodicity trajectories of DNA sequences are not smooth, but with many noises. The nonlinear TD is used to smooth the trajectories and get the derivatives of the signals.

1. Smooth the 3-base periodicity trajectories by nonlinear TD and compute the derivatives. If the derivative at the position is larger than 0, set the nucleotide at the position as exon nucleotide; otherwise, set it as intron nucleotide.

2. Reduce false exon and false intron. There are only few exon or intron of length less than 50 bp. Thus, for a predicted exon of length less than 50 bp, the region is considered false positive; similarly, if a predicated intron less than 50 bp, the region is considered as false negative.

### E. The performance evaluation of gene finding algorithm

The performance of gene finding method is measured in terms of sensitivity, specificity and accuracy, which are defined as follows [15]. The sensitivity $S_n = TP/(TP+FN)$ and the specificity $S_p = TN/(TN+FP)$, where *TP* is the true positive, which is the length of nucleotides of correctly predicted exons; *TN* is the true negative, which is the length of nucleotides of correctly predicted introns; *FN* is the false negative, which is the length of nucleotides of wrongly predicted introns; and *FP* is the false positive, which is the length of nucleotides of wrongly predicted exons. In other words, *Sn* is the proportion of coding sequences that have been correctly predicted as coding, and *Sp* is the proportion of non-coding sequences that have been correctly predicted as non-coding. The accuracy *AC* is defined as the average of *Sn* and *Sp*.

## III. RESULTS AND DISCUSSION

It was shown that an exon displays an upward trend in the plot of the 3-base periodicity per base pair, while an intron displays a flat trend [14]. Thus, for a given DNA sequence, putative exons and introns can be identified from the derivatives of the 3-base periodicity per base pair along the DNA sequence [4]. Fig. 2(a) the plot of the 3-base periodicity per base pair of the gene at locus *AAB26989.1* of *Drosophila melanogaster (fruit fly)*. There are approximate changing points between exons and introns in the plot, and the plot are with a lot of noise. Fig. 2(b) and 2(c) are the plots of the 3-base periodicity per base pair of the gene after the data was filtered by the nonlinear TD with the gain values as 0.01 and 0.001, respectively. Fig. 2(b) and Fig. 2(c) shows the smoothness of the 3-base periodicity trajectories of DNA sequences by the nonlinear TD, indicating the nonlinear TD is an effective approach to approximate data to obtain derivatives.

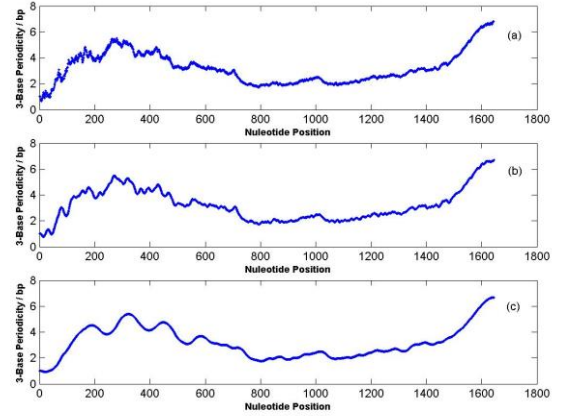

Figure 2. Tracking of signal to noise of Fourier spectrum of gene. (a) Original data, (b) R = 0.01, (c) R = 0.001 by Nonlinear TD with R = 0.5.

To evaluate the performance of the algorithm in protein coding region prediction, the method is applied to predict the intron and exon of structure known genes from different organisms. As an example, Figure 3(a) is expected structure of the gene at locus *AAB26989.1* of *Drosophila melanogaster (fruit fly)*. Figure 3(b) is the predicted gene structures using the nonlinear TD with the gain values of 0.001. Fig. 3(c) is the predicted gene structure after removing the false intron and false exon. The values of *Sn, Sp* and *AC* of the prediction in Figure 3(c) are 0.9763, 0.5992 and 0.7877, respectively. Using the original method without TD smoothing described in [14], the values of *Sn, Sp* and *AC* of the prediction are 0.9169, 0.5908, and 0.7539, respectively. Optimizing the gene prediction method can be obtained through smoothing the 3-base periodicity trajectories at different levels at different gain values. For example, the *AC* values of the prediction using the gain values 0.1 and 0.0001 are 0.5537 and 0.7192, respectively. Both of them are lower than the one at the gain value of 0.001. This result indicates TD smoothing method using the gain value of 0.001 may improve the gene prediction accuracy about 3.5%.

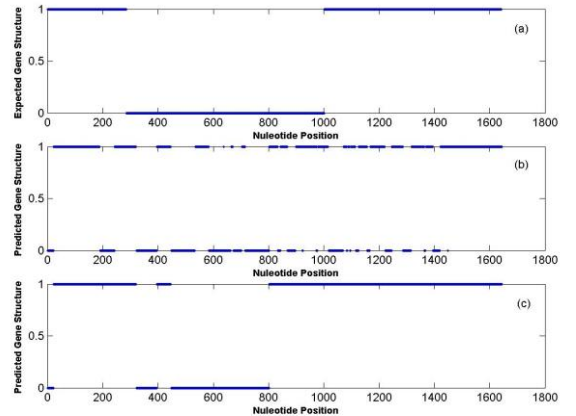

Figure 3. Gene structure based on tracking the 3-base periodicity using the nonlinear TD. (a) Expected gene structure. (b) Predicted gene structure (R = 0.001 for the nonlinear TD). (c) Predicated gene structure after removing the false introns and false exons.

Many experiment data have local noise due to high sensitivity of measurement methods and may be corrupted by the noise. It is important to remove noise to enhance the signal to noise level before performing data analysis. There are several methods, such as Savitzky-Golay filter, polynomial regressing, discrete Wavelet transform, etc, to smooth signals with noise [16], but these methods may not get derivatives of the smoothed data directly and have other limitations. The derivatives of the signal changes are important to understand many mechanisms in biological systems, for example, the dynamics of the signals [17] and gene finding [14]. Among these smoothing, nonlinear TD is valuable for data processing for its capacity to find a smooth signal which converges to the original signal and obtain the derivative at the same time. It is easy to apply without considering of the complexity for the noised signal. In addition, using the exceeding factor R provides a fine tuning approach at different approximation levels.

ACKNOWLEDGMENT

This research is supported by the U.S. NSF grant (DMS-1120824) and National Natural Sciences Foundation of China (31271408).

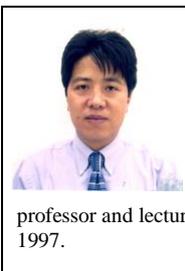
**Changchuan Yin,** Ph.D., is with Faculty of Mathematics, College of Natural Sciences, University of Phoenix Chicago Campus, IL, USA. He received Ph.D.(2005) in mathematics and M.S.(2001) in mathematical computer sciences from University of Illinois at Chicago, USA, and M.S.(1993) in biochemistry from Fudan University, Shanghai, China. He was assistant professor and lecturer of biochemistry of Fudan University from 1993 to 1997.

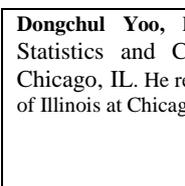
**Dongchul Yoo,** Ph.D., is with Department of Mathematics, Statistics and Computer Science, University of Illinois at Chicago, IL. He received Ph.D.(2005) in mathematics from University of Illinois at Chicago

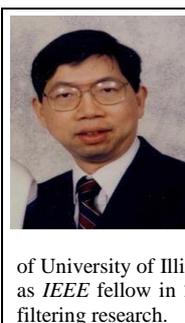
**Stephen S.-T. Yau,** Ph.D., Professor, is with Department of Mathematical Sciences, Tsinghua University, Beijing, China. He received Ph.D.(1976) and M.A.(1974) from State University of New York at Stony Brook (USA). He was awarded as Distinguished Professor of Department of Mathematics, Statistics and Computer Science of University of Illinois at Chicago (USA) in 2005. He was director of Control and Information Laboratory of University of Illinois at Chicago from 1993 to 2011. He was awarded as *IEEE* fellow in 2003 for his distinguished contribution to nonlinear filtering research.